\documentclass[12pt]{article}

\usepackage{graphics}
\usepackage{amssymb}

\newcommand{\e}{\mathrm{e}}

\newcommand{\C}{\mathbb{C}}

\newcommand{\R}{\mathbb{R}}

\newtheorem{claim}{Claim}[section]
\newtheorem{theorem}[claim]{Theorem}
\newtheorem{lemma}[claim]{Lemma}
\newtheorem{conjecture}[claim]{Conjecture}
\newtheorem{remark}[claim]{Remark}

\newenvironment{proof}[1][Proof]{\textsl{#1.} }{\ \rule{0.4em}{0.7em}}

\begin{document}

\title{Scattering by local deformations \\ of a straight leaky wire}
\author{P.~Exner$^{a,b}$ and S.~Kondej$^c$}
\date{}
\maketitle

\begin{quote}
{\small \em a) Nuclear Physics Institute, Academy of Sciences,
25068 \v Re\v z \\ \phantom{a) }near Prague, Czech Republic
\\
b) Doppler Institute, Czech Technical University,
B\v{r}ehov{\'a}~7, \\ \phantom{a) }11519 Prague, Czech Republic
\\
c) Institute of Physics, University of Zielona G\'{o}ra, ul.
Szafrana 4a, \\ \phantom{a) } 65246 Zielona G\'{o}ra, Poland
\\
\phantom{a) }\texttt{exner@ujf.cas.cz},
\texttt{skondej@if.uz.zgora.pl} }
\end{quote}

\begin{quote}
{\small We consider a model of a leaky quantum wire with the
Hamiltonian $-\Delta -\alpha \delta(x-\Gamma)$ in $L^2(\R^2)$,
where $\Gamma$ is a compact deformation of a straight line. The
existence of wave operators is proven and the S-matrix is found
for the negative part of the spectrum. Moreover, we conjecture
that the scattering at negative energies becomes asymptotically
purely one-dimensional, being determined by the local geometry in
the leading order, if $\Gamma$ is a smooth curve and $\alpha
\to\infty$.}
\end{quote}


\section{Introduction}
\label{introd} \setcounter{equation}{0}

Graph models are very useful in describing a variety of mesoscopic
systems -- see \cite{Ku} for a review. They have some drawbacks,
however, namely that they contain free parameters in boundary
conditions describing the vertices, with no easy way to fix their
values, and they neglect quantum tunnelling between different
parts of the graph. An attempt to construct models free of these
deficiencies was a motivation of the recent work on \emph{leaky
quantum graphs} described formally by Schr\"odinger operators
\begin{equation}\label{formex}
  -\Delta -\alpha \delta(x-\Gamma)
\end{equation}
in $L^2(\R^d)$ with an attractive singular interaction supported
by a graph $\Gamma$; a precise definition of this operator,
denoted as $-\Delta _{\Gamma}$ or $-\Delta _{\alpha,\Gamma}$ will
be given below for the particular situation considered in this
paper.

Various results are available concerning the discrete spectrum of
such systems -- see, e.g., \cite{BT, BEKS}, more recently
\cite{EI, EK1, EK2, EY1, EY2, EY3} and references given in these
papers. On the other hand, almost nothing is known about the
scattering in this context, apart from analysis of a very simple
model \cite{EK3} and indirect indications coming from spectral
properties; both indicated that interesting resonance effects may
occur \cite{EN}.

Our aim in this paper is to address this question in the simple
situation when the system is planar, $d=2$, and $\Gamma$ is a
local deformation of a straight line $\Sigma =\{(x_1
,0):\,x_{1}\in \R\}$, or in other words, that the perturbation of
$-\Delta_\Sigma= -\Delta -\alpha \delta(x-\Sigma)$ responsible for
the scattering is a singular interaction supported by the
symmetric difference of the two sets,
 \begin{equation}\label{symmdiff}
 \Lambda\equiv \Gamma\triangle\Sigma:=(\Gamma \setminus \Sigma )
 \cup (\Sigma \setminus \Gamma)\,.
 \end{equation}
The spectrum of the unperturbed operator $-\Delta_\Sigma$ is
easily found by separation of variables. The transverse part is
reduced to the problem to one-dimensional Laplacian with a single
point interaction \cite{AGHH}. Since the latter is attractive
gives rise to the eigenvalue $-\alpha ^2 /4$ with the
eigenfunction $\mathrm{e}^{-\alpha |x|/2}$. Consequently, the
two-dimensional system described by $-\Delta_\Sigma$ has a purely
absolutely continuous spectrum equal to $(-\alpha^2 /4,\infty)$;
states with negative energies can only be transported along the
line $\Sigma$.

With the singular character of the perturbation in mind our main
tool will be a Krein-type resolvent formula which we derive on
Sec.~\ref{Kreinres} below. The assumption about compact support
will allow us to check stability of the essential spectrum in
Sec.~\ref{secspecrum}, and moreover, to derive the same result for
the absolutely continuous spectrum and to prove the existence of
the wave operators -- see Sec.~\ref{waveop}. The spectral
properties of $-\Delta_\Sigma$ suggest that the scattering problem
looks differently for positive and negative energies; on this
paper we concentrate on the \emph{negative spectrum} of
$-\Delta_\Gamma $. The generalized eigenfunctions of the
unperturbed operator $-\Delta_\Sigma $ corresponding to
eigenvalues $\lambda \in (-\alpha^2 /4, 0)$ are easily seen to be
\begin{equation}\label{gensigma}
\omega_{\lambda}(x_1 ,x_2 )=\mathrm{e}^{i(\lambda +\alpha^2 /4
)^{1/2}x_{1}}\mathrm{e}^{-\alpha|x_{2}|/2}
\end{equation}
and its complex conjugate $\bar\omega_{\lambda}$. The generalized
eigenfunctions of $-\Delta_\Gamma $ will be in Sec.~\ref{smatrix}
constructed as superpositions of $\omega_{\lambda}$ and
$\bar\omega_{\lambda}$ when we are far from the scattering region
$\Lambda$, so that the scattering problem is essentially
one-dimensional in the sense that it is described by a $2\times 2$
matrix of reflection and transmission amplitudes. The scattering
problem for the positive part of spectrum is more complicated and
we postpone it to a subsequent paper.

The claim about one-dimensional character has to be taken
\emph{cum grano salis} because due to quantum tunnelling the
scattering depends in general on the global geometry of $\Gamma$
as an example worked out in \cite{EN} suggests. One can expect,
however, that such effects will be suppressed if the attractive
interaction is strong enough. In the concluding remarks we will
make this claim more precise stating it as a conjecture which is
expected to be valid in the asymptotic regime $\alpha\to\infty$,
in the leading order at least, if $\Gamma$ is a sufficiently
smooth curve.


\section{Scattering due to local deformation}
\label{sec:scatt} \setcounter{equation}{0}

\subsection{Geometry of $\Gamma$ and definition of $-\Delta _{\Gamma}$}

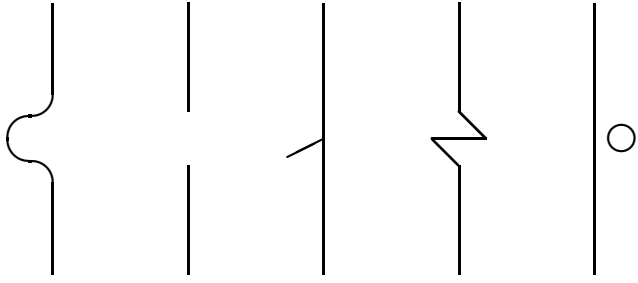
\begin{figure} \label{defline}
 \setlength\unitlength{1.2mm}
 \begin{picture}(95,0)(30,55)
 \thicklines
 \put(55,40){\line(0,1){10}}
 \put(55,60){\line(0,1){10}}
 \put(52.5,50){\oval(5,5)[tr]}
 \put(52.5,60){\oval(5,5)[br]}
 \put(52.5,55){\oval(5,5)[l]}
 \put(70,40){\line(0,1){12}}
 \put(70,58){\line(0,1){12}}
 \put(85,40){\line(0,1){30}}
 \put(85,55){\line(-2,-1){4}}
 \put(100,40){\line(0,1){12}}
 \put(100,58){\line(0,1){12}}
 \put(97,55){\line(1,0){6}}
 \put(100,52){\line(-1,1){3}}
 \put(100,58){\line(1,-1){3}}
 \put(115,40){\line(0,1){30}}
 \put(118,55){\circle{3}}
 \end{picture}
 \vspace{20mm}
 \caption{Examples of locally deformed lines}
 \end{figure}

Naturally we have to assume more about $\Gamma$ than just its
local character; we suppose that $\Gamma$ is a finite family of
$C^1$ smooth curves in $\R^2$. We will also require that no pair
of components of $\Gamma$ crosses in their interior points,
neither a component has a self-intersection; we allow the
components to touch at their endpoints but assume they do not form
a cusp there. To summarize this survey of requirements we assume
that
  \begin{description}
  \item{\bf{(a1)}}
there exists a compact set $M\subset\R^2$ such that
\begin{equation}\label{assum1}
\Gamma \setminus M= \Sigma \setminus M \,,
\end{equation}
\item{\bf{(a2)}} the set $\Gamma \setminus \Sigma $ admits a
finite decomposition,
\begin{equation}\label{decomG}
\Gamma \setminus \Sigma = \bigcup_{i=1}^{N}\Gamma _{i}\,,\quad
N<\infty\,,
\end{equation}
where the $\Gamma _{i}$'s are finite $C^{1}$ curves with the
properties described above.
\end{description}
Examples of such locally deformed lines are shown in
Fig.~\ref{defline}. More comments on the assumptions will be given
below -- cf.~Remark~\ref{remarka}.

Let us described next a proper way to define the Hamiltonian with
a perturbation supported by $\Gamma$. For $i=1,...,N$ we denote by
$\nu_{i}$ the Dirac measure on $\Gamma_{i}$, more precisely, for a
Borel set $\mathcal{B}\subset \R^2$ we have
$$
\nu_i (\mathcal{B}):=l(\mathcal{B} \cap \Gamma_{i} )\,,
$$
where $l(\cdot )$ is the one-dimensional Hausdorff measure given
by the arc length of $\Gamma_i$; in a similar way we define the
measure $\tilde{\nu }$ on $\Sigma \cap \Gamma$. Then the sum
$\eta:=\tilde{\nu}+\sum_{i=1}^{N}\nu_{i}$ is a Dirac measure on
$\Gamma$ and it follows from Theorem~4.1 of \cite{BEKS} that it
belongs to the generalized Kato class. We also introduce the space
$L^{2}(\eta )\equiv L^{2}(\R^{2}, \eta )$ which admits the direct
sum decomposition
\begin{equation}\label{dirsumL}
L^{2}(\eta )=L^{2}(\tilde{\nu})\oplus \left ( \bigoplus _{i=1}^{N}
L^{2}(\nu_{i})\right )\,.
\end{equation}
A rigorous definition of $-\Delta _{\Gamma }$ can be in terms of
the following quadratic form,
\begin{equation}\label{form}
  \gamma (f,g):=(\nabla f,\nabla g)-\alpha (I_{\Gamma }f,I_{\Gamma
  }g)_{L^{2}(\eta )}\quad \mathrm{for}\quad  f,g\in W^{2,1}
  \equiv W^{2,1}(\R^{2})\,,
\end{equation}
where $(\cdot, \cdot)$ is the scalar product in $L^2\equiv
L^2(\R^2 )$ and $I_{\Gamma }$ is the standard embedding operator
acting from $W^{2,1}$ to $L^{2}(\eta)$. For brevity we will write
in the following $(f,g)_{L^{2}(\eta)}= (I_{\Gamma}f,
I_{\Gamma}g)_{L^{2}(\eta)}$ assuming that the functions $f,g\in
W^{2,1}$ are embedded in $L^{2}(\eta)$, and the same
self-explanatory notations will be used for other spaces with
Dirac measures. Since the measure $\eta $ is of the Kato class we
infer that the form (\ref{form}) is closed \cite{BEKS}, and
therefore the operator associated with it is self-adjoint; we
identify it with the Hamiltonian of the problem given formally by
(\ref{formex}).

\begin{remark} \label{remarka}
{\rm The assumptions can be slightly weakened. For instance, one
can require only that the components $\Gamma_i$ are only
\emph{piecewise} $C^1$ smooth, which is equivalent to gluing
several of them into a single curve. Since the corresponding
$\eta$ has to belong to the generalized Kato class cusps must be
avoided. To formulate a sufficient condition for that let us
parameterize $\Gamma_i$ by its arc length, i.e. regard it as a
graph of the function $(0,L)\ni s\mapsto \Gamma _{i}(s)\in \R^2$.
Cusps will be then absent if there is a $C_i>0$ such that
\begin{equation}\label{assa2'}
|\Gamma _{i}(s)-\Gamma _{i}(s')|\geq C_i |s-s'|\quad
\mathrm{for}\quad s\,,s'\in (0,L)\,.
\end{equation}
The same condition can be used for $\Gamma$ with branching points;
one has to take all possible piecewise smooth curves which are
subsets of such a $\Gamma$ and to demand that they satisfy the
above inequality.}
\end{remark}

\subsection{Krein type formula for the resolvent
of $-\Delta _{\Gamma }$} \label{Kreinres}

Since $-\Delta_\Gamma$ is a singular perturbation of $-\Delta$ one
can write the corresponding relation between the resolvents  --
cf.~\cite{EI}. For our purposes, however, it is more useful to
regard it as a singular perturbation of $-\Delta_{\Sigma}$ by a
$\delta$ potential supported by the set $\Lambda$ which will
decompose as follows,
\begin{equation}\label{Lambda}
\Lambda =\Lambda_0 \cup \Lambda _1 \quad \mathrm{with}\quad
\Lambda_{0}:=\Sigma \setminus \Gamma \,,\,\,\, \Lambda_{1}:=\Gamma
  \setminus \Sigma = \bigcup_{i=1}^{N}\Gamma _{i}\,;
\end{equation}
the coupling constant of the potential representing the
perturbation will be naturally $\alpha $ on $\Lambda_{0}$ and
$-\alpha $ on $\Lambda_{1}$.

Recall first how the resolvent of $-\Delta_{\Sigma}$ looks like.
Assume that  $\mathrm{Im}\,k>0$ and $k^{2}$ belongs to the
resolvent set of the free Laplacian, $k^2 \in \rho (-\Delta )$,
and denote by $R^k$ the resolvent of $-\Delta $, which is an
integral operator with the kernel
$$
G^{k}(x\!-\!y)=\frac{1}{(2\pi)^{2}}\int_{\R^2}
\frac{\mathrm{e}^{ip(x-y)}}{p^{2}-k^{2}}\,\mathrm{d}p=
\frac{1}{2\pi}K_{0}(ik(x\!-\!y))\,,
$$
where $K_0 (\cdot)$ stands for the Macdonald function. To
construct resolvent of $-\Delta _{\Sigma}$ we need the embeddings
of $R^k$ to spaces canonically associated with $\Sigma$. Let
$\mu_{\Sigma }\equiv \mu$ be the Dirac measure on $\Sigma$; by
means of it we define the operator
$$
\mathrm{R}^{k}_{\mu}\,:\,L^{2}(\mu)\rightarrow L^{2}\,,\quad
\mathrm{R}^{k}_{\mu}f=G^{k}\ast f\mu
$$
with the adjoint $(\mathrm{R}^{k}_{\mu})^{*}\,:L^{2}\rightarrow
L^{2}(\mu)$ and $\mathrm{R}^{k}_{\mu \mu}$ which is the integral
operator with the same kernel as $\mathrm{R}^{k}_{\mu}$ but acting
from $L^{2}( \mu)$ to $L^{2}( \mu)$. Using the natural isomorphism
$L^{2}(\mu)\cong L^{2}(\R)$ the kernel of $\mathrm{R}^{k}_{\mu
\mu}$ can be written as
\begin{equation}\label{kerGmm}
G_{\mu \mu}^{k}(x_{1}-y_{1})=\frac{1}{4
\pi}\int_{\R}\frac{\mathrm{e}^{ip_{1}(x_{1}-y_{1})}}{\tau
_{k}(p_{1})}\,\mathrm{d}p_{1}\,,
\end{equation}
where $\tau _{k}(p_{1}) :=(p^{2}_{1}-k^{2}) ^{1/2}$. Given $k^2
\in \rho (-\Delta _{\Sigma})=\C\setminus (-\alpha^2 /4, \infty )$
with $\mathrm{Im}\,k>0$ we can express the resolvent of $-\Delta
_\Sigma$ in the following form
$$
R_{\Sigma }^{k}=R^{k}+\alpha \mathrm{R}^{k}_{\mu}(1-\alpha
\mathrm{R}^{k}_{\mu \mu} )^{-1}(\mathrm{R}^{k}_{\mu})^{*}\,,
$$
which is, of course, a particular case of a general formula given
in \cite{EI}. A straightforward calculation using (\ref{kerGmm})
shows that $R_{\Sigma }^{k}$ is an integral operator with the
kernel
\begin{equation}\label{kernelsig}
G^{k}_{\Sigma}(x\!-\!y)=G^{k}(x\!-\!y)+\frac{\alpha}{4\pi^{3}}
\int_{\R^{3}}\frac{\mathrm{e}^{ipx-ip'y}}{(p^{2}\!-\!k^{2})
(p'^{2}\!-\!k^{2})} \frac{\tau _{k}(p_{1})}{2\tau
_{k}(p_{1})\!-\!\alpha}\, \mathrm{d}p\, \mathrm{d}p'_{2}\,,
\end{equation}
where we have denoted $p=(p_{1},p_{2})$ and $p'=(p_{1},p'_{2})$.

Now we are going to express the resolvent of $-\Delta_\Gamma$
understanding this operator as a singular perturbation of $-\Delta
_\Sigma $. We want to derive a Krein-type formula using $R_{\Sigma
}^{k}$ and its appropriate embeddings to $L^2 (\nu )$, where $\nu
\equiv \nu_{\Lambda}$ is the Dirac measure on $\Lambda$. The
latter allows for the following decomposition
\begin{equation}\label{decompnu}
\nu =\nu_\Lambda =\nu_{0}+\sum_{i=1}^{N}\nu_{i}\,,
\end{equation}
where $\nu_0$ is the Dirac measure on $\Lambda_0$. It convenient
also to denote $\mathrm{h}\equiv L^{2}(\nu)$; this space inherits
from (\ref{decompnu}) the direct sum decomposition
$\mathrm{h}=\mathrm{h}_{0}\oplus \mathrm{h}_{1}$ with
$\mathrm{h}_{0}\equiv L^{2}(\nu_{0})$ and $\mathrm{h}_{1}\equiv
\bigoplus_{i=1}^{N}L^{2}(\nu_{i})$. In the same way as before we
introduce the operator
 \begin{equation} \label{tracop}
\mathrm{R}_{\Sigma ,\nu}^{k}:\mathrm{h}\rightarrow L^2 \,,\quad
\mathrm{R}_{\Sigma ,\nu}^{k}f=G^{k}_{\Sigma}\ast f\nu\quad
\mathrm{for}\quad f\in \mathrm{h}
 \end{equation}
which will be shown to be defined on the whole $\mathrm{h}$ for
suitable values of $k$. Similarly, $(\mathrm{R}_{\Sigma
,\nu}^{k})^{\ast}:L^2 \rightarrow \mathrm{h}$ is its adjoint and
$\mathrm{R}_{\Sigma ,\nu \nu}^{k}$ denotes the operator-valued
matrix in $\mathrm{h}$ with the ``block elements'' $G^{k}_{\Sigma
,ij} \equiv G^{k}_{\Sigma ,_{\nu_{i}\nu_{j}}}:L^{2}(\nu_{j})\to
L^{2}(\nu_{i})$ defined as the appropriated embeddings of
(\ref{kernelsig}). The following lemmata show that the above
constructed operators are bounded at least for some $k$.

\begin{lemma} \label{boundG}
For any $a>0$ there exists $\kappa_a$ such that the inequality
$$
\|G^{i\kappa }\ast f\nu_i \|_{L^2 (\nu_{j})}\leq a
\|f\|_{L^{2}(\nu_i )}\,,\quad f\in L^{2}(\nu_i )
$$
holds for all $\kappa >\kappa _{a}$.
\end{lemma}
\begin{proof} The argument is the same as in Corollary~2.2 of \cite{BEKS}.
\end{proof}

\begin{lemma} \label{proposi1}
\noindent \emph{(i)} For any $\kappa \in (\alpha/2, \infty)$ the
operator $\mathrm{R}_{\Sigma ,\nu}^{i\kappa}$ is bounded. \\
\noindent \emph{(ii) } For any $\sigma >0$ there exists
$\kappa_\sigma $ such that for $\kappa > \kappa _\sigma$ the
operator $\mathrm{R}_{\Sigma ,\nu \nu}^{i\kappa}$ is bounded with
the norm less than $\sigma $.
\end{lemma}
\begin{proof}
(i) It suffices to establish the existence of $C>0$ such that
\begin{equation}\label{proposi1a}
\|G _{\Sigma }^{k}\ast f\nu_{j} \|\leq C \|f
\|_{L^{2}(\nu_{j})}\,,\quad f\in L^{2}(\nu_{j}) \,,
\end{equation}
holds for $j=0,...,N$ and $k=i\kappa$, where $\kappa \in
(\alpha/2, \infty)$. One can consider the terms at the r.h.s. of
(\ref{kernelsig}) separately. For the first component $G^k$ of
$G^{k}_{\Sigma }$ it follows from Sobolev embedding theorem. Let
us denote the second component of $G^{k}_{\Sigma }$ in
(\ref{kernelsig}) by $\xi^k$. For any $\kappa \in (\alpha /2 ,
\infty)$ we have the inequality
\begin{equation}\label{esttau}
0< \frac{\tau_{i\kappa }(p_1 )}{2\tau_{i\kappa }(p_1 )-\alpha
}<M_\kappa \quad \mathrm{for}\quad p_1\in\mathbb{R}
\end{equation}
with a constant $M_\kappa>0$.  Hence an elementary estimate gives
\begin{equation}\label{proposi1ab}
\|\xi^{k}\ast f\nu_{j}\|^{2}\leq C_{1}\|f\|^{2}_{L^{1}(\nu_{j})}
\int_{\R^{2}}\frac{1}{(p^{2}+\kappa ^{2})^{2}(p_{1}^{2}+\kappa ^2
)}\,\mathrm{d}p\leq C_{2}\|f\|^{2}_{L^{2}(\nu_{j})}\,,
\end{equation}
where $C_1\,,C_2$ are positive constants and $p=(p_1 ,p_2 )$; in
the last inequality we have used the fact that $\nu_j$ has a
compact support. This yields (\ref{proposi1a}).

\noindent (ii) Another straightforward estimate relying on
(\ref{esttau}) yields
$$
\|\xi^{k}\ast f\nu_{j}\|^{2}_{L^{2}(\nu_{j})} \leq
C_{1}'\|f\|^{2}_{L^{1}(\nu_{j})} \left( \int_{\R
}\frac{1}{p^{2}_{1}+\kappa ^{2}}\right )^{2}\mathrm{d}p_1 \leq
C_{2}'\kappa ^{-2}\|f\|^{2}_{L^{2}(\nu_{j})}\,,
$$
for each $i,j=0,...,N$, where $C_{1}'\,,C_{2}'$ are positive
constants. In view of (\ref{esttau}) we see that $C_{1}'\,,C_{2}'$
are in fact functions of $\kappa$ but they are uniformly bounded
w.r.t. $\kappa >\kappa _0$ where $\kappa _0 \in (\alpha /2 ,\infty
)$ is a fixed number. Combining this result with
Lemma~\ref{boundG} we arrive at the desired conclusion.
\end{proof}

\begin{remark}\label{remembedding}
\rm{ Note that $\mathrm{R}_{\Sigma ,\nu}^{i\kappa}$ with $\kappa
\in (\alpha /2,\infty)$ is in fact a continuous embedding to
$W^{2,1}$. Indeed, the Sobolev space theory tells us that
$$
\|G^k \ast f\nu_{j} \|_{W^{2,1}} <C\|f \|_{L^{2}(\nu_{j})}\,,\quad
f\in L^{2}(\nu_{j})\,.
$$
On the other hand, the estimate (\ref{proposi1ab}) can be
strengthened,
$$
\|\xi^{k}\ast f\nu_{i}\|^{2}_{W^{2,1}}\leq
C_{1}\|f\|^{2}_{L^{1}(\nu_{i})}
\int_{\R^{2}}\frac{1}{(p^{2}+\kappa ^{2})(p_{1}^{2}+\kappa ^2
)}\,\mathrm{d}p\leq C_{2}\kappa ^{-2}\|f\|^{2}_{L^{2}(\nu_{i})}\,;
$$
together these results give the above claim. }
\end{remark}

\smallskip

\noindent Let us now introduce an operator-valued matrix acting in
$\mathrm{h}=\mathrm{h}_{0}\oplus\mathrm{h}_{1}$ as
$$
\Theta ^{k}=-(\alpha
^{-1}\check{\mathbb{I}}+\mathrm{R}^{k}_{\Sigma ,\nu\nu})  \quad
\mathrm{with} \quad \check{\mathbb{I}}= \left(
\begin{array}{cc}\mathbb{I}_{0}
& 0   \\
0 & -\mathbb{I}_{1}  \end{array} \right)\,,
$$
where $\mathbb{I}_{i}$ are the unit operators in $\mathrm{h}_{i}$.
By Lemma~\ref{proposi1} the operator $\Theta^{i\kappa}$ is
boundedly invertible if $\kappa $ is large enough, i.e.
$(\Theta^{i\kappa} )^{-1}\in \mathcal{B}(\mathrm{h})$. Now we are
ready to prove the following theorem.

\begin{theorem}\label{}
Let $(\Theta ^{k})^{-1}\in \mathcal{B}(\mathrm{h})$ hold for $k\in
\C^+$ and let the operator
\begin{equation}\label{resolvent}
  R^{k}_{\Gamma }=R^{k}_{\Sigma }+\mathrm{R}^{k}_{\Sigma ,\nu}
  (\Theta^{k})^{-1}(\mathrm{R}^{k}_{\Sigma ,\nu})^{*}
\end{equation}
be defined everywhere on $L^2$. Then $k^{2}$ belongs to $\rho
(-\Delta _{\Gamma })$ and the resolvent $(-\Delta
_\Gamma-k^2)^{-1}$ is given by $R^{k}_{\Gamma }$.
\end{theorem}
\begin{proof}
Notice first that in view of Lemma~\ref{proposi1} the assertion is
not empty; let us suppose for the moment that $k=i\kappa $ with
$\kappa $ sufficiently large. Furthermore, the quadratic form
(\ref{form}) can be rewritten as follows,
\begin{equation}\label{formau}
\gamma (f,g)=(\nabla f,\nabla g)- \alpha
(f,g)_{L^{2}(\mu_{\Sigma})}+\alpha
(\check{\mathbb{I}}I_{\Lambda}f,I_{\Lambda}g)_{\mathrm{h}}\,,\quad
f,\,g\in W^{2,1}\,,
\end{equation}
where $I_{\Lambda}$ is the standard embedding of $W^{2,1}$ to
$\mathrm{h}=L^{2}(\nu_{\Lambda})$. By Remark~\ref{remembedding} we
have $f=R^{k}_{\Gamma}h\in W^{2,1}$ for $h\in L^2$, and applying
(\ref{resolvent}) to (\ref{formau}) we get
\begin{equation}\label{resocal}
\gamma (f,g)-k^{2}(f,g)
=(h,g)+((\Theta^{k})^{-1}(\mathrm{R}^{k}_{\Sigma
,\nu})^{*}h,I_{\Lambda}g)_{\mathrm{h}} +\alpha
(\check{\mathbb{I}}I_{\Lambda}R^{k}_{\Gamma
}h,I_{\Lambda}g)_{\mathrm{h}}\,.
\end{equation}
To proceed further let us note that the definitions of
$I_{\Lambda}$ and $\Theta^{k}$ imply
\begin{eqnarray}\label{resocal1}
\lefteqn{ \alpha (\check{\mathbb{I}}I_{\Lambda
}R^{k}_{\Sigma}h,I_{\Lambda}g)_{\mathrm{h}} =\alpha
(\check{\mathbb{I}}(\mathrm{R}^{k}_{\Sigma,\nu
})^{\ast}h,I_{\Lambda}g)_{\mathrm{h}}} \nonumber \\ && =
-((\Theta^{k })^{-1}(\mathrm{R}^{k}_{\Sigma
})^{\ast}h,I_{\Lambda}g)_{\mathrm{h}}- \alpha
(\check{\mathbb{I}}\mathrm{R}^{k}_{\Sigma ,\nu
\nu}(\Theta^{k})^{-1}(\mathrm{R}^{k}_{\Sigma ,\nu
})^{*}h,I_{\Lambda}g)_{\mathrm{h}}\,. \phantom{AAA}
\end{eqnarray}
Applying again (\ref{resolvent}) to (\ref{resocal}) and using
(\ref{resocal1}) we get by a direct calculation that $\gamma
(f,g)-k^{2}(f,g)=(h,g)$ holds for any $g\in W^{2,1}$. This is
equivalent to the relation $R^{k}_{\Gamma }=(-\Delta
_{\Gamma}-k^{2})^{-1}$ for $k=i\kappa $ with $\kappa $ is
sufficiently large, and as the resolvent of a self-adjoint
operator $R^{k}_{\Gamma }$ can continued analytically to the
region $\C^+$; this completes the proof.
\end{proof}

\subsection{Spectrum of $-\Delta _\Gamma$}  \label{secspecrum}

Let us turn to the description of the spectrum of our Hamiltonian.
The spectrum of the unperturbed operator $-\Delta_\Sigma$ is found
easily by separation of variables. The transverse part is the
one-dimensional operator $-\Delta _{\alpha}^{(1)}$ with a single
point interaction. It is well known \cite{AGHH} that its spectrum
is purely absolutely continuous in $[0,\infty)$, and in the
attractive case, $\alpha>0$, which we are interested in, there is
also one eigenvalue equal to $-\frac{1}{4}\alpha ^2$. Combining
this with the Laplacian in the other direction, we find that the
spectrum of $-\Delta_\Sigma$ is purely absolutely continuous
covering the interval $[-\frac{1}{4}\alpha ^2, \infty)$.

Let us first check stability of the essential spectrum.

\begin{theorem} \label{essspec}
$ \sigma _{\mathrm{ess}}(-\Delta _\Gamma )=\sigma
_{\mathrm{ess}}(-\Delta _\Sigma )=\left[-\frac{1}{4}\alpha ^2,
\infty \right)\,. $
\end{theorem}
\begin{proof}
In view of the resolvent formula (\ref{resolvent}) and the Weyl
theorem it is sufficient to show that there exists $k\in \C^+$
such that the operator
$$
B^{k}\equiv \mathrm{R}^{k}_{\Sigma ,\nu}
(\Theta^{k})^{-1}(\mathrm{R}^{k}_{\Sigma ,\nu})^{*}\,
$$
is compact. It follows from Lemma~\ref{proposi1} that
$(\Theta^{i\kappa})^{-1}\in \mathcal{B}(\mathrm{h})$ and
$(\mathrm{R}^{i\kappa}_{\Sigma ,\nu})^{*}$ is bounded if $\kappa$
is large enough. Furthermore, it was shown in \cite{BEKS} that
\begin{equation}\label{aux1}
\int_{\R^2}\int_{\R^2}|G^{i\kappa}(x\!-\!y)|^{2}
\,\nu_{j}(\mathrm{d}y)\,\mathrm{d}x<\infty\,.
\end{equation}
On the other hand for $\kappa \in (\alpha /2 ,\infty)$ and
$j=0,...,N$ the second component of $G^{i\kappa}_\Sigma$ given in
(\ref{kernelsig}) can be estimated
\begin{equation}\label{aux2}
\int_{\R^2}\int_{\R^2}|\xi^{k}(x, y)|^{2}
\,\nu_{j}(\mathrm{d}y)\,\mathrm{d}x<CL_{j}
\int_{\R^2}\frac{\mathrm{d}p}{(p^{2}+\kappa)^{2}}<\infty\,,
\end{equation}
where $C$ is a constant and $L_j$ denote the length of
$\Lambda_j$. Combining (\ref{aux1}) and (\ref{aux2}) we get the
compactness of $\mathrm{R}^{k}_{\Sigma ,\nu}$ , and thus the same
for $B^{k}$.
\end{proof}

 \begin{remark}
 {\rm Using the results of \cite{AP2} we can determine the discrete
 spectrum of $-\Delta_{\Gamma }$ from zeros of the operator-valued
 function $k\mapsto \Theta ^k$. It is known, for example, that
 if $\Gamma $ is a single non-straight curve there is at least one
 isolated eigenvalue below $-\frac{1}{4}\alpha ^2\;$ \cite{EI}.
 However, the discrete spectrum is not the object of our interest in
 the present paper. }
 \end{remark}

\subsection{Existence of wave operators}\label{waveop}

Let us turn now to the scattering theory for the pair $(-\Delta
_{\Gamma },-\Delta _{\Sigma })$. To establish the existence of
wave operators we will employ the Kuroda--Birman theorem. This is
made possible by the following result.
\begin{theorem}
$B^{i\kappa}$ is a trace class operator for $\kappa$ sufficiently
large.
\end{theorem}
\begin{proof}
The idea is borrowed in part from \cite{BT}. By
Lemma~\ref{proposi1} we have
$$
(\Theta ^{i\kappa} )^{-1}\leq C'(\Theta ^{i\kappa,+})^{-1},\;\;
\mathrm{where }\;\; \Theta^{i\kappa, +}:=\alpha
^{-1}\mathbb{I}+\mathrm{R}_{\Sigma ,\nu \nu }^{i\kappa}\;\;
\mathrm{and}\;\; \mathbb{I}= \left(
\begin{array}{cc}\mathbb{I}_{0}
& 0   \\
0 & \mathbb{I}_{1}  \end{array} \right),
$$
for some $C'>0$ and all $\kappa $ sufficiently large; it is clear
that operator $(\Theta ^{i\kappa,+})^{-1}$ is positive and
bounded. This in turn implies the inequality
$$
B^{i\kappa}\leq C' B^{i\kappa, +}\,,\quad B^{i\kappa, +}:=
\mathrm{R}^{i\kappa}_{\Sigma ,\nu}
(\Theta^{i\kappa,+})^{-1}(\mathrm{R}^{i\kappa}_{\Sigma
,\nu})^{*}\,.
$$
Furthermore, define $B^{i\kappa, +}_\delta $ as the integral
operator with the kernel
$$
B^{i\kappa,+}_ \delta (x,y)=\chi
_{\delta}(x)B^{i\kappa,+}(x,y)\chi _{\delta} (y)\,,
$$
where $B^{i\kappa, +}(\cdot ,\cdot)$ is the kernel of $B^{i\kappa,
+}$ and $\chi _{\delta}$ stands for the indicator function of the
ball $\mathcal{B}(0,\delta )$; one has, of course, $B_{ \delta
}^{i\kappa,+}\to B^{i\kappa,+}$ as $\delta \to \infty$ in the weak
sense. Moreover, using the estimate from the proof of
Theorem~\ref{essspec} we get
\begin{eqnarray*}
\lefteqn{\int_{\R^{2}}B^{i\kappa,+}_ \delta (x,x)\mathrm{d} x =
\int_{\R^{2}}(G^{i\kappa}_{\Sigma }(\cdot ,x)\chi _{\delta}
(x),(\Theta ^{i\kappa, +})^{-1}G^{i\kappa}_{\Sigma }(\cdot ,x)
\chi_{\delta}(x))_{\mathrm{h}}\,\mathrm{d}x } \\
 && \le \| (\Theta^{i\kappa,+})^{-1}\| \int _{\R^2} \|
 G_{\Sigma }^{i\kappa} (\cdot ,x)
 \chi_{\delta}(x)\|^{2}_{\mathrm{h}}\,\mathrm{d}x \leq C
\nonumber \|(\Theta ^{i\kappa, +})^{-1}\|\,, \phantom{AAA}
\end{eqnarray*}
where $C$ is a positive constant. Next we apply the lemma
following Thm~XI.31 in \cite{RS} by which the operator
$B^{i\kappa, +}_\delta$ is trace class for any $\delta>0$ (and
$\kappa$ large enough); since $\mathrm{Tr\,} B^{i\kappa, +}_\delta
\to \mathrm{Tr\,} B^{i\kappa, +}$ holds as $\delta \to \infty$,
the same is true also for the limiting operator. In a similar way
we can construct a Hermitian trace class operator $B^{i\kappa,-}$
which provides an estimate from below, $B^{i\kappa,-}\leq
B^{i\kappa}$; this means that $B^{i\kappa}$ is a trace class too.
\end{proof}

\section{Generalized eigenfunctions and the \\ S-matrix}
\label{smatrix} \setcounter{equation}{0}

The existence of wave operators itself does not tell us much, we
have to be able to find the S-matrix, which by definition acts as
$$
S\psi_{\lambda }^{-}=\psi_{\lambda }^{+}
$$
relating the incoming and outgoing asymptotic solutions. In
particular, for scattering in the negative part of the spectrum
with a fixed $\lambda \in (-\frac{1}{4}\alpha ^2,0)$ corresponding
to the effective momentum $k_{\alpha }(\lambda ):=(\lambda +\alpha
^{2}/4)^{1/2}$, the latter are combinations of the generalized
eigenfunctions $\omega_{\lambda}$ and $\bar\omega_{\lambda}$ given
by (\ref{gensigma}).

The functions (\ref{gensigma}) and their analogues $\omega_z$ for
complex values of the energy parameter are $L^2$ only locally, of
course, and we can approximate them by the family of regularized
functions,
$$
\omega _{z}^{\delta }(x)=\mathrm{e}^{-\delta
x_{1}^{2}}\omega_{z}(x)\quad \mathrm{for} \quad z\in \rho (-\Delta
_{\Sigma })\,,
$$
which naturally belong to $D(-\Delta _{\Sigma })$. Consider now a
function $\psi _{z}^{\delta }$ such that $(-\Delta _{\Gamma
}-z)\psi _{z}^{\delta }=(-\Delta _{\Sigma }-z)\omega _{z}^{\delta
}$. A direct computation gives
\begin{equation}\label{psiconv}
(-\Delta _{\Gamma  }-z)\psi _{z}^{\delta }=2\delta (2\delta
x_{1}^{2}-1-2i k_{\alpha}(\lambda ))\psi _{z}^{\delta }\,.
\end{equation}
After taking the limit $\lim_{\epsilon \to 0}\psi_{\lambda
+i\epsilon }^{\delta }=\psi_{\lambda }^{\delta }$ in the topology
of $L^2$ the function $\psi_{\lambda }^{\delta }$ still belongs to
$D(-\Delta _{\Gamma })$, and moreover, we have
$$
\psi_{\lambda }^{\delta }=\omega _{\lambda }^{\delta
}+\mathrm{R}_{\Sigma ,\nu}^{k_\alpha(\lambda )}
(\Theta^{k_\alpha(\lambda )})^{-1}I_{\Lambda} \omega_{\lambda
}^{\delta }\,,
$$
where $\mathrm{R}_{\Sigma ,\nu}^{k_\alpha(\lambda )}$ is the
integral operator acting on the auxiliary Hilbert space
$\mathrm{h}$, analogous to (\ref{tracop}), which is given by the
kernel
\begin{equation}\label{limitspec}
G^{k_\alpha(\lambda )}_{\Sigma }(x\!-\!y):=\lim_{\varepsilon \to
0}G^{k_\alpha(\lambda +i\varepsilon)}_{\Sigma}(x\!-\!y)\,;
\end{equation}
similarly $\Theta^{k_\alpha(\lambda )}:= -\alpha
^{-1}\check{\mathbb{I}}-\mathrm{R}^{k_\alpha(\lambda )}_{\Sigma
,\nu\nu}$ are the operators on $\mathrm{h}$ with
$\mathrm{R}^{k_\alpha(\lambda )}_{\Sigma ,\nu\nu}$ being the
embeddings defined by means of (\ref{limitspec}). The explicit
form of this kernel was derived in \cite{EK2} to be\footnote{To be
fully specific, the formula is obtained from eq.~(4.8) of
\cite{EK2} after interchanging $x_1 \to x_1 - y_1$ and $a \to
y_2$.}
\begin{eqnarray}
\lefteqn{G^{k_\alpha(\lambda )}_{\Sigma }(x\!-\!y)=
K_{0}(i\sqrt{\lambda }(x\!-\!y))} \nonumber \\&& +\mathcal{P} \int
_{0}^{\infty} \frac{\mu_{0}(t;x,y)}{t-\lambda - \alpha ^{2}/4}\,
\mathrm{d}t+ s_{\alpha}(\lambda)\, \mathrm{e}^{ik_{\alpha
}(\lambda)|x_{1}-y_{1}|}\,\e^{-\alpha /2 (|x_{2}|+|y_{2}|)},
\end{eqnarray}
where $s_{\alpha}(\lambda):=i\alpha (2^{3}k_{\alpha
}(\lambda))^{-1}$ and
$$
\mu_{0}(t;x,y):=-\frac{i\alpha }{2^5
\pi}\,\frac{\mathrm{e}^{it^{1/2}(x_1 -y_1
)}\,\mathrm{e}^{-(t-\lambda)^{1/2}(|x_2 |+|y_2
|)^{1/2}}}{t^{1/2}((t-\lambda)^{1/2})}\,.
$$
Of course, the pointwise limits $\psi _{\lambda }=\lim_{\delta \to
0 }\psi _{\lambda }^{\delta }$ cease to be square integrable,
however, they still belongs locally to $L^2$, and in view of
(\ref{psiconv}) they provide us with the generalized eigenfunction
of $-\Delta _{\Gamma }$ in the form
\begin{equation}\label{geneg2}
\psi_{\lambda }=\omega_{\lambda }+\mathrm{R}_{\Sigma
,\nu}^{k_\alpha(\lambda )} (\Theta^{k_\alpha(\lambda
)})^{-1}J_{\Lambda} \omega_{\lambda }\,,
\end{equation}
where $J_{\Lambda} \omega_{\lambda }$ is an embedding of
$\omega_{\lambda }$ to $L^{2}(\nu_{\Lambda})$.
 To find the S-matrix we have to investigate the behavior of
$\psi _{\lambda } $ for  $|x_{1}|\to \infty$. We employ the
following result.

\begin{lemma}
Let $y$ belong to a compact $M\subset \R^{3}$ and $|x_{1}|\to
\infty$, then
$$
G^{k_\alpha(\lambda )}_{\Sigma }(x\!-\!y)\approx s_{\alpha
}(\lambda )\, \mathrm{e}^{ik_{\alpha }(\lambda)|x_{1}-y_{1}|}\,
e^{-\alpha /2 (|x_{2}|+|y_{2}|)}\,.
$$
\end{lemma}
\begin{proof}
The argument is the same as in \cite{EK2}.
\end{proof}

\bigskip

This allows us to formulate the sought conclusion.
\begin{theorem} \label{geneigenvTH}
For a fixed $\lambda \in (-\frac{1}{4}\alpha^2,0)$ the generalized
eigenfunctions behave asymptotically as
\begin{equation}\label{psibeh2}
\psi_\lambda(x) \approx \left\{ \begin{array}{lcl}
\mathcal{T}(\lambda )\,\mathrm{e}^{ik_{\alpha }(\lambda)x_{1}} \,
\mathrm{e}^{-\alpha|x_{2}|/2} & \;\; \mathrm{for } \;\; & x_{1}\to
\infty \\ [.3em]
\mathrm{e}^{ik_{\alpha}(\lambda)x_{1}}\mathrm{e}^{-\alpha|x_{2}|/2}
+ \mathcal{R}(\lambda )\,\mathrm{e}^{-ik_{\alpha
}(\lambda)x_{1}}\mathrm{e}^{-\alpha|x_{2}|/2} & \;\; \mathrm{for
}\;\; & x_{1}\to -\infty \end{array} \right.
\end{equation}
where $k_{\alpha }(\lambda ):=(\lambda +\alpha ^{2}/4)^{1/2}$ and
$\mathcal{T}(\lambda )\,,\mathcal{R}(\lambda )$ are the
transmission and reflection amplitudes given respectively by
$$
\mathcal{T}(\lambda )=1-s_{\alpha}(\lambda )
\left((\Theta^{k_\alpha(\lambda )})^{-1}J_\Lambda \omega _{\lambda
},J_\Lambda \omega_{\lambda }\right)_{\mathrm{h}}\nonumber
$$
and
$$
\mathcal{R}(\lambda )=s_{\alpha}(\lambda )
\left((\Theta^{k_\alpha(\lambda )})^{-1} J_\Lambda \omega
_{\lambda },J_\Lambda \bar{\omega} _{\lambda
}\right)_{\mathrm{h}}\,.
$$
\end{theorem}


\section{Concluding remarks}\label{concl}
\setcounter{equation}{0}

The general formulae for the S-matrix coefficients given in the
Theorem~\ref{geneigenvTH} are not easy to handle and one would
like to ask whether there are situations when there is a simple
way, at least in a perturbative sense. Let us explain in more
details how such a result could look like, assuming that $\Gamma$
is a $C^{4}$ smooth curve obtained by a local deformation of a
straight line (the conditions (a1), (a2) are then, of course,
fulfilled) and $\alpha$ is large enough. We expect that
$\mathcal{T}, \,\mathcal{R}$ will be in this situation expressed
in the leading order through the \emph{local} geometry of
$\Gamma$.

We may suppose without loss of generality that the curve is
parameterized by its arc length being the graph of a function
$$\R\ni s\mapsto (\Gamma^{(1)}(s),\Gamma^{(2)}(s))\in\R\, .$$
By assumption the curvature $\kappa (\cdot )$ of $\Gamma$ is well
defined and allows us to define a \emph{comparison operator}, the
same as in \cite{EY1}, by
$$
K:D(K)\rightarrow L^{2}(\R )\,,\quad
K=-\frac{d^{2}}{ds^{2}}-\frac{1}{4}\kappa ^{2}(s)\,,
$$
with the natural domain $D(K):=W^{2,2}(\R)$. It is nothing else
than a one-dimensional Schr\"odinger operator with an attractive
compactly supported $C^2$ smooth potential. The corresponding
scattering problem is thus well posed and we denote by
$\mathcal{T}_{K}(k),\, \mathcal{R}_{K}(k)$ the corresponding
transmission and reflection amplitudes at a fixed momentum $k$.
Denote by $\mathbf{S}_{\Gamma ,\alpha}(\lambda)$ and
$\mathbf{S}_{K}(\lambda)$ the on-shell $S-$matrixes of
$-\Delta_\Gamma$ and $K$ at energy $\lambda$, respectively. Then
we can make the following conjecture about the asymptotic
behaviour.

\begin{conjecture} \label{srong-coup}
For a fixed $k\ne 0$ and $\alpha\to\infty$ we have the relation
\begin{equation}\label{S-mtri-a}
\mathbf{S}_{\Gamma ,\alpha}\Big(k^2- \frac{1}{4} \alpha^2\Big)\to
\mathbf{S}_{K}(k ^{2})\,.
\end{equation}
\end{conjecture}
The claim is inspired by the corresponding result about the
discrete spectrum of such systems \cite{EY1, EY2} which uses
natural curvilinear coordinates in the vicinity of the curve to
express the solution to the Schr\"odinger equation through that of
the comparison problem plus an error term which vanishes as
$\alpha\to\infty$. In the present case, however, one cannot use
bracketing and minimax estimates and has to investigate instead
directly the generalized eigenfunction in a strip neighbourhood of
$\Gamma$; it is sufficient to find the behaviour of the solution
in the straight asymptotic parts where $|x_1|$ is large. We
postpone this analysis to a later publication.

Another open question which the considerations given in this paper
raise is whether our results, notably Theorem~\ref{geneigenvTH},
extend to more general situations when $\Gamma$ is no longer a
local perturbation of a straight line but it remains to be
asymptotically straight in a suitable sense. We expect that the
answer will be positive, however, one will need to replace wave
operators of Sec.~\ref{waveop} by generalized ones referring to
the asymptotes of $\Gamma$ with an appropriate identification map.

\bigskip

\noindent \emph{The research has been partially supported by ASCR
within the project K1010104 and by the Polish Ministry of
Scientific Research and Information Technology under (solicited)
grant No PBZ-Min-008/PO3/03.}


\end{document}